\pdfoutput=1
\documentclass[aps,prb,reprint,floatfix]{revtex4-1}
\usepackage{amsmath,amsfonts,amssymb} %
\usepackage{mathtools} %
\usepackage{wasysym} %
\usepackage{bm} %
\usepackage[bbgreekl]{mathbbol} %
\usepackage{graphicx} %
\usepackage[dvipsnames,table]{xcolor} %
\usepackage{tabularx} %
\usepackage[acronym]{glossaries} %
\usepackage{braket} %
\usepackage[binary-units=true,detect-weight=true]{siunitx} %
\usepackage{fancyhdr} %
\usepackage{microtype} %
\usepackage{natmove}
\usepackage[caption=false]{subfig}
\usepackage[byname]{smartref} %
\PassOptionsToPackage{obeyspaces}{url}
\usepackage[breaklinks, %
colorlinks, linkcolor=Blue, citecolor=Blue, urlcolor=Blue]{hyperref} %
\usepackage[version=3]{mhchem}
\AtBeginDocument{\usepackage{booktabs}} 

\newcommand{\bfr}{\mathbf{r}} %
\newcommand{\E}{\textrm{e}} %
\newcommand{\I}{\mathrm{i}\mkern1mu} %
\newcommand{\RR}{\mathbf{R}} %
\newcommand{\rr}{\mathbf{r}} %
\def\be{\begin{equation}} %
\def\ee{\end{equation}} %
\def\bea{\begin{eqnarray}} %
\def\eea{\end{eqnarray}} %

\DeclareSIUnit[number-unit-product = {\,}]\cal{cal}

\def\MyTitle{Quantum chemistry on quantum annealers} %
\def\MyAuthora{Scott N. Genin} %
\def\MyAuthorb{Ilya G. Ryabinkin} %
\def\MyAuthorc{Artur F. Izmaylov} %
\def\MySubject{Quantum computing} %

\hypersetup{ %
  pdftitle={\MyTitle{}}, %
  pdfauthor={\MyAuthora{}, \MyAuthorb{}, and \MyAuthorc}, %
  pdfsubject={\MySubject{}}, %
  pdfkeywords={quantum chemistry, qubit coupled clusters, annealers,
    global optimization} %
}

\newcolumntype{Y}{>{\centering\arraybackslash}X}

\newacronym{JW}{JW}{Jordan--Wigner} %
\newacronym{BK}{BK}{Bravyi--Kitaev} %
\newacronym{QPE}{QPE}{quantum phase estimation} %
\newacronym{VQE}{VQE}{variational quantum eigensolver} %
\newacronym{UCC}{UCC}{unitary coupled cluster} %
\newacronym{QMF}{QMF}{qubit mean-field} %
\newacronym{QCC}{QCC}{qubit coupled cluster} %
\newacronym{CI}{CI}{configuration interaction} %
\newacronym{FCI}{FCI}{full configurational interaction} %
\newacronym{CASCI}{CASCI}{complete active space configurational
  interaction} %
\newacronym{MCSCF}{MCSCF}{multiconfigurational self-consistent
  field} %
\newacronym{CAS}{CAS}{complete active space} %
\newacronym{CASSCF}{CASSCF}{complete active space self-consistent
  field} %
\newacronym{CCSD}{CCSD}{coupled-cluster singles and doubles} %
\newacronym{RHF}{RHF}{restricted Hartree--Fock} %
\newacronym{UHF}{UHF}{unrestricted Hartree--Fock} %
\newacronym{PES}{PES}{potential energy surface} %
\newacronym{PEC}{PEC}{potential energy curve} %
\newacronym[longplural={degrees of freedom}, firstplural={degrees of
  freedom (DOF)}, plural={DOF}]{DOF}{DOF}{degree of freedom} %
\newacronym[longplural={equations of motion}, firstplural={equations
  of motion (EOM)}, plural={EOM}]{EOM}{EOM}{equation of motion} %
\newacronym{SQP}{SQP}{sequential quadratic programming} %
\newacronym{MMA}{MMA}{method of moving asymptotes} %

\begin{document}

\title{\MyTitle}

\author{\MyAuthora{}} %
\affiliation{OTI Lumionics Inc., 100 College St. \#351, Toronto,
  Ontario\, M5G 1L5, Canada} %

\author{\MyAuthorb{}} %
\email{ilya.ryabinkin@otilumionics.com}
\affiliation{OTI Lumionics Inc., 100 College St. \#351, Toronto,
  Ontario\, M5G 1L5, Canada} %
\affiliation{Department of Physical and Environmental Sciences,
  University of Toronto Scarborough, Toronto, Ontario\, M1C\,1A4,
  Canada} \affiliation{Chemical Physics Theory Group, Department of
  Chemistry, University of Toronto, Toronto, Ontario\, M5S 3H6,
  Canada}

\author{\MyAuthorc{}} %
\affiliation{Department of Physical and Environmental Sciences,
  University of Toronto Scarborough, Toronto, Ontario\, M1C\,1A4,
  Canada} \affiliation{Chemical Physics Theory Group, Department of
  Chemistry, University of Toronto, Toronto, Ontario\, M5S 3H6,
  Canada}

\date{\today}

\begin{abstract}
  Quantum chemistry calculations for small molecules on quantum
  hardware have been demonstrated to date only on universal-gate
  quantum computers, not quantum annealers. The latter devices are
  limited to finding the lowest eigenstate of the Ising Hamiltonian
  whereas the electronic Hamiltonian could not be mapped to the Ising
  form without exponential growth of the Ising Hamiltonian with the
  size of the system
  [\href{http://dx.doi.org/10.1021/acs.jpcb.7b10371}{J. Phys. Chem. B
    \textbf{122}, 3384 (2018)}]. Here we propose a novel mixed
  \emph{discrete-continuous} optimization algorithm, which finds the
  lowest eigenstate of the \gls{QCC} method using a quantum annealer
  for solving a discrete part of the problem. The \gls{QCC} method is
  a potentially exact approach for constructing the electronic
  wavefunction in the qubit space. Therefore, our methodology allows
  for systematically improvable quantum chemistry calculations using
  quantum annealears. We illustrate capabilities of our approach by
  calculating \gls{QCC} ground electronic states for the LiH,
  \ce{H2O}, and \ce{C6H6} molecules. \ce{C6H6} calculations involve 36
  qubits and are the largest quantum chemistry calculations made on a
  quantum annealer (the D-Wave 2000Q system) to date. Our findings
  opens up a new perspective for use quantum annealers in
  high-throughput material discovery.
\end{abstract}

\glsresetall

\maketitle

\section{Introduction}

Quantum chemistry simulations are often considered as an ideal
application of quantum computers following inspiring ideas of
R.~Feynman~\cite{Feynman:1982/ijtp/467}. This tacitly assumes that the
quantum computer is the universal one, capable of simulating quantum
evolution governed by an arbitrary Hamiltonian. However, fundamental
and technological obstacles with building such a universal quantum
device prompted researchers and engineers to consider more limited
architectures, such as quantum annealers. Quantum
annealers~\cite{Johnson:2011/nature/194, Boixo:2014/natphys/218,
  Lechner:2015/scia/e1500838}, also known as Ising
machines~\cite{Inagaki:2016/sci/603, McMahon:2016/sci/614,
  Inagaki:2016/natpt/415}, can only find the ground state of the Ising
Hamiltonian~\cite{Kadowaki:1998/pre/5355},
\begin{equation}
  \label{eq:Ising_H}
  \hat H_\text{Is} =  \sum_{i=1}^N h_i \hat z_i + \sum_{i,j = 1}^N
  J_{ij} \hat z_i \hat  z_j,
\end{equation}
where $\hat z_i$ is the Pauli $z$-operator acting on the $i$-th spin
(qubit), $h_i$ and $J_{ij}$ are constants
that can be tuned independently~\cite{Johnson:2011/nature/194}.

Since the molecular Hamiltonian is not in the Ising form, the
electronic structure problem cannot be set up and solved directly on
quantum annealers. However, there is still a strong impetus for use
annealers due to their intrinsic ability to solve hard optimization
problems~\cite{Santoro:2002/sci/2427} (but also see a counterexample,
Ref.~\citenum{Battaglia:2005/pre/066707}). To date, only one
work~\cite{Xia:2018/jpcb/3384} proposed a mapping of a general qubit
Hamiltonian to the Ising form. Unfortunately, the qubit size of the
resulting Ising Hamiltonian grows exponentially with the size of the
system making this approach viable only for small systems (\emph{e.g.}
\ce{H2} and \ce{LiH}). Moreover, to reach chemical accuracy
($\le \SI{1}{\kilo\cal\per\mol}$) for \ce{H2} in the minimal STO-3G
basis required $\sim 1400$ qubits on the D-Wave 2000Q
system~\cite{Streif:2018/arXiv/1811.05256}.

Here we take a different route. Instead of trying to use quantum
annealing for the whole problem, we employ it as a part of a hybrid
quantum-classical scheme that accelerates the convergence in the
\gls{QCC} method~\cite{Ryabinkin:2018/jctc/6317}. This method has been
originally introduced for solving the electronic structure problem on
a universal quantum computer within the \gls{VQE}
framework~\cite{Peruzzo:2014/ncomm/4213, Wecker:2015/pra/042303}.
Here, we do not employ a universal quantum computer but instead use
the \gls{QCC} energy functional for establishing the variational
optimization problem. Even though this optimization problem is
nonlinear, certain symmetries of the \gls{QCC} energy functional
allows us to substantially reduce the domain of continuous
optimization variables by introducing auxiliary discrete variables.
Discrete optimization is usually an exponentially difficult problem
requiring combinatorial search, and it may seem that such a reduction
of the domain only makes the problem harder. However, in this case, it
is possible to perform the discrete optimization by finding the lowest
eigenstate of some Ising Hamiltonian. Therefore, a quantum annealer
becomes essential in reducing the complexity of the \gls{QCC}
nonlinear optimization by solving the discrete part of the problem.

The rest of the paper is organized as follows. After a brief review of
the electronic structure problem and the \gls{QCC} method we show how
the domain reduction idea can be integrated into the QCC framework. In
particular, we discuss how the discrete optimization can be introduced
in the \gls{QCC} formalism, and how the Ising Hamiltonian whose lowest
eigenstate is the solution for the discrete problem can be formulated.
We illustrate our developments by solving the electronic structure
problem for \ce{LiH}, \ce{H2O}, and \ce{C6H6} molecules on a simulated
perfect quantum annealer and the D-Wave 2000Q system\cite{dwave2000q}.

\section{Theory}
\label{sec:theory}

\subsection{Electronic structure problem}
\label{sec:electr-struct-probl}

Electronic structure calculations amount to finding the solution of
the time-independent electronic Schr\"odinger equation,
\begin{equation}
  \label{eq:el_se}
  \hat H_e \Psi_i(\bar\rr | \bar\RR) = E_i(\bar\RR) \Psi_i(\bar\rr |
  \bar\RR).
\end{equation}
Here $\hat H_e$ is the electronic Hamiltonian of a molecule with
electronic variables
$\bar\rr = (\mathbf{r}_1, \dots, \mathbf{r}_{N_e})$ and nuclear
configuration parameters
$\bar\RR = (\mathbf{R}_1, \dots, \mathbf{R}_N)$. $E_i(\bar\RR)$ and
$\Psi_i(\bar\rr | \bar\RR)$ are \glspl{PES} and electronic wave
functions, respectively. Eq.~\eqref{eq:el_se} is a many-body fermionic
problem that defines electronic properties of molecules and materials
from first principles, \emph{i.e.} solely from knowledge of type and
location of nuclei and the number of electrons $N_e$.

For transforming Eq.~\eqref{eq:el_se} to a qubit form, the
differential operator $\hat H_e$ is considered as an operator in a
finite-dimensional Fock space using the second quantization formalism:
\begin{equation}
  \label{eq:qe_ham}
  \hat H_e = \sum_{ij} h_{ij} {\hat a}^\dagger_i {\hat a}_j + \frac{1}{2}\sum_{ijkl}
  \Braket{ij|kl} {\hat a}^\dagger_i {\hat a}^\dagger_j {\hat a}_l {\hat a}_k.
\end{equation}
Here ${\hat a_i}^\dagger$ (${\hat a_i}$) are fermionic creation
(annihilation) operators, and
\begin{align}
  \label{eq:one-ints}
  h_{ij} & = \int \psi_i^*(\mathbf{x}) \left(-\frac{1}{2} \nabla_{\bfr}^2 - \sum_\alpha \frac{Z_\alpha}{|\bfr -
           \mathbf{R}_\alpha|}\right)
           \psi_j(\mathbf{x})\,\mathrm{d}\mathbf{x}, \\
  \label{eq:two-ints}
  \Braket{ij|kl} & = \int \psi_i^*(\mathbf{x}_1)
                   \psi_j^*(\mathbf{x}_2) \frac{1}{r_{12}}
                   \psi_k(\mathbf{x}_1)
                   \psi_k(\mathbf{x}_1)\,
                   \mathrm{d}\mathbf{x}_1\mathrm{d}\mathbf{x}_2
\end{align}
are one- and two-electrons integrals, respectively.
$\{\psi_i(\mathbf{x})\}_{i=1}^{N_\text{so}}$ are the spin-orbitals,
which depend on a joined (spatial plus spin) coordinate of an
electron, $\mathbf{x} = (\bfr, \sigma)$, and constitute a spin-orbital
basis of the size $N_\text{so}$. Typically spin-orbitals are
themselves constructed as linear expansions over an auxiliary basis
set of atomic-centered functions known as atomic orbitals.

The size of the one-electron basis determines the size of the matrix
representation of $\hat H_e$, which is
$2^{N_\text{so}} \times 2^{N_\text{so}}$. Thus, the exact algebraic
solution is possible for molecules containing only few atoms.
Eigenvectors of an operator~\eqref{eq:qe_ham} are known as \gls{FCI}
states. Corresponding eigen-energies are commonly used as benchmarks
for any approximate methods as they can be only improved by enlarging
the one-electron basis set.

Using one of the conventional fermion-to-qubit transformations, such
as the \gls{JW}~\cite{Jordan:1928/zphys/631,
  AspuruGuzik:2005/sci/1704} or \gls{BK}~\cite{Bravyi:2002/aph/210,
  Seeley:2012/jcp/224109, Tranter:2015/ijqc/1431,
  Setia:2017/ArXiv/1712.00446, Havlicek:2017/pra/032332}, the
second-quantized fermionic Hamiltonian~\eqref{eq:qe_ham} can be
iso-spectrally transformed to a qubit form,
\begin{equation}
  \label{eq:spin_ham}
  \hat H = \sum_I C_I\,\hat T_I,
\end{equation}
where $C_I$ are deduced from one- and two-electron integrals ($h_{ij}$
and $\Braket{ij|kl}$), and $\hat T_I$ operators are products of
several spin operators,
\begin{equation}
  \label{eq:Ti}
  \hat T_I = \cdots \hat \sigma_1^{(I)} \hat  \sigma_0^{(I)}, 
\end{equation}
which we call ``Pauli words'' for brevity. Each of
$\hat \sigma_i^{(I)}$, is one of the Pauli $\hat x_i$, $\hat y_i$, or
$\hat z_i$ operators.

\subsection{\Acrlong{QCC} method}
\label{sec:brief-descr-qmf}

The \gls{QCC} method relies on a two-tier parametrization of a trial
wave function: 1) the \gls{QMF}
description~\cite{Ryabinkin:2019/jctc/249, Ryabinkin:2018/jcp/214105}
and 2) multi-qubit transformations to account for electron
correlation~\cite{Ryabinkin:2018/jctc/6317}. The \gls{QMF} part uses
the simplest variational Ansatz that is possible on a quantum
computer: a direct product of superposition states of individual
qubits,
\begin{equation}
  \label{eq:qcc_wf}
  \ket{\boldsymbol \Omega} = \prod_{i=1}^{N_q} \ket{\Omega_i},
\end{equation}
where
\begin{equation}
  \label{eq:coh_state}
  \ket{\Omega_i} = \cos\left(\frac{\theta_i}{2}\right) \ket{\alpha}
  + \E^{\I\phi_i}\sin\left(\frac{\theta_i}{2}\right) \ket{\beta}
\end{equation}
is a so-called spin-coherent state for the $i$-th
qubit~\cite{Radcliffe:1971/jpa/313, Arecchi:1972/pra/2211,
  Perelomov:1972, Lieb:1973/cmp/327}. $\phi_i$ and $\theta_i$ are
azimuthal and polar angles on the ``Bloch sphere'' of the $i$-th
qubit, respectively, and $\ket{\alpha}$ and $\ket{\beta}$ are ``up''
and ``down'' eigenstates of the $\hat z_i$ operator. The \gls{QMF}
ground-state energy is defined as a minimum of the corresponding
energy functional with respect to all Bloch angles
$\boldsymbol \Omega = \{\phi_i, \theta_i\}_{i=1}^{N_q}$:
\begin{equation}
  \label{eq:emf}
  E_\text{QMF} = \min_{\boldsymbol \Omega} \braket{\boldsymbol
    \Omega|\hat H|\boldsymbol \Omega}.
\end{equation}

The energy functional~\eqref{eq:emf} has an exceptionally simple form
in terms of Bloch angles. To derive it, one needs to replace all Pauli
operators in Eq.~\eqref{eq:spin_ham} with functions according to the
rule
\begin{equation}
  \label{eq:qmf_rule}
  \begin{array}{ccl}
    \hat x_i & \to & \cos\phi_i\sin\theta_i, \\
    \hat y_i & \to & \sin\phi_i\sin\theta_i, \\
    \hat z_i & \to & \cos\theta_i
  \end{array}
\end{equation}
and convert operator products to ordinary products of real numbers.
The domain of definition for angles is
\begin{align}
  \label{eq:phi_domain}
  \phi_i & \in [0,\ 2\pi), \\
  \label{eq:theta_domain}
  \theta_i & \in [0,\ \pi), \quad i = 1, \dots, N_q.
\end{align}
In what follows we consider the \gls{QMF} energy function as a
separate approximation to the solution of the electronic structure
problem.

The second step in the \gls{QCC} method introduces a multi-qubit
unitary transformation
\begin{equation}
  \label{eq:Ucc}
  U(\boldsymbol \tau) = \prod_{k=1}^{N_\text{ent}} \exp(-\I \tau_k \hat P_k/2), 
\end{equation}
where $\hat P_k$ are the multi-qubit Pauli words (``entanglers''),
which are responsible for multi-qubit entanglement, and $\tau_k$ are
the corresponding amplitudes that are optimized within a domain
\begin{equation}
  \label{eq:ampl_domain}
  \tau_k \in [0, 2\pi),\ k = 1, \dots, N_\text{ent}.
\end{equation}
The total \gls{QCC} energy assumes the form
\begin{equation}
  \label{eq:qcc_form}
  E_\text{QCC} = \min_{\boldsymbol \Omega, \boldsymbol \tau} \braket{\boldsymbol
    \Omega| U^\dagger(\boldsymbol \tau) \hat H U(\boldsymbol \tau)|\boldsymbol \Omega}.
\end{equation}

The transformed Hamiltonian
$U^\dagger(\boldsymbol \tau) \hat H U(\boldsymbol \tau)$ in
Eq.~\eqref{eq:qcc_form} can be calculated recursively by the
formula~\cite{Ryabinkin:2018/jctc/6317}:
\begin{align}
  \label{eq:qcc_func_single_tau}
  \hat A^{(k)}(\tau_k, \dots, \tau_1) = {}
  & \E^{\I \tau_k \hat P_k/2}\,\hat
    A^{(k-1)}(\tau_{k-1}, \dots, \tau_1) \, \E^{-\I \tau_k \hat
    P_k/2} \nonumber \\ 
  = {}
  & \hat A^{(k-1)} - \I \frac{\sin \tau_k}{2} [\hat A^{(k-1)}, \hat P_k] \nonumber \\
  & +  \frac{1}{2}\left(1 - \cos \tau_k\right) \hat P_k\, [\hat A^{(k-1)}, \hat P_k],
\end{align}
where $k = 1, \dots, N_\text{ent}$ and $\hat A^{(0)} = \hat H$. This
procedure produces $3^{N_\text{ent}}$ distinct operator terms, but
frequently good results can be achieved already at small
$N_\text{ent}$. The problem of optimal choice of entanglers is
addressed in Ref.~\citenum{Ryabinkin:2018/jctc/6317}, and we assume
here that it is already solved, so that
Eq.~\eqref{eq:qcc_func_single_tau} has been used $N_\text{ent}$ times
to generate a list of operators and trigonometric factors that depend
on $\{\tau_k\}$. The final form of the \gls{QCC} energy functional can
now be obtained by applying the rule~\eqref{eq:qmf_rule} to each of
the operators in the list and summing them together. The resulting
expression is a function of amplitudes $\boldsymbol \tau$ and angles
$\boldsymbol \Omega$. Classical minimization of that function yields
the \gls{QCC} ground-state energy.

The \gls{QMF} and \gls{QCC} energy functions are sums of products,
where each individual term consists of the factors $\sin(\phi_i)$,
$\cos(\phi_i)$, $\sin(\theta_i)$, $\cos(\theta_i)$, $\sin(\tau_i)$,
and $[1 - \cos(\tau_i)]$ that occur no more than once; in other words,
they are polylinear functions of those factors. As our experience
shows, the search for the global minimum of \gls{QMF} or \gls{QCC}
energy starting from a random guess is a difficult task; the
minimization procedure tends to converge to different local minima.
The situation is very much like as in the conventional \gls{MCSCF}
method:~\cite{Celestino:2003/mp/1937} the corresponding non-linear
equations have multiple solutions. Note that such a problem is less
common (albeit possible) in the single-configuration Hartree--Fock
method: in the most of the implementations the Fock matrix and its
eigenvalues---orbital energies---are avaliable, and one can enforce
the Aufbau principle by populating the orbitals with the lowest
energies first~\cite{Saunders:1973/ijqc/699}, avoiding high-energy
local minima that describe core-hole or highly excited Rydberg states.
Unfortunately, this option is not available in the \gls{QMF} method.
Thus, a strategy how to maximize the likelihood of finding the global
minimum is needed.

\subsection{Domain reduction by folding}
\label{sec:domain-folding}

\begin{figure*}
  \centering
  \subfloat[][]{\includegraphics[width=0.33\textwidth]{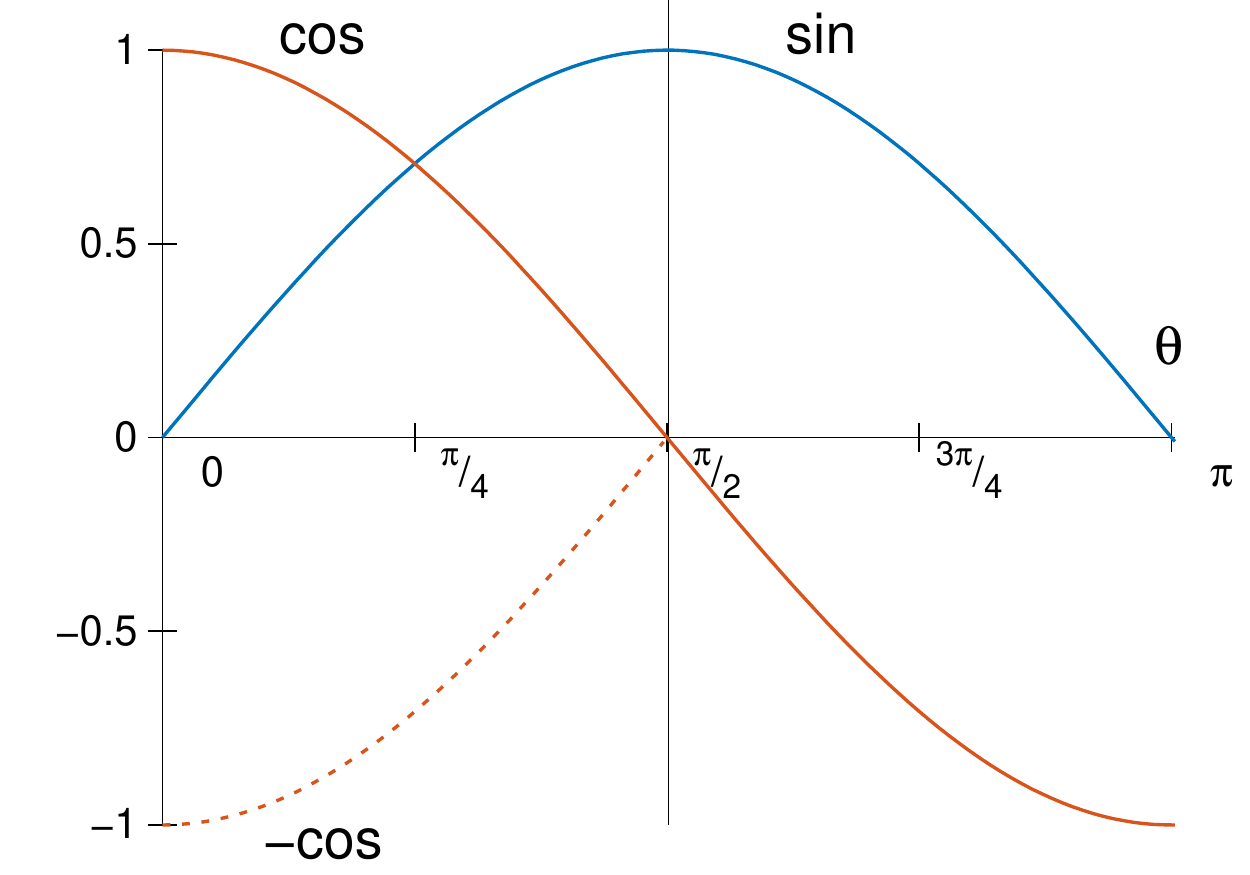}}
  \subfloat[][]{\includegraphics[width=0.33\textwidth]{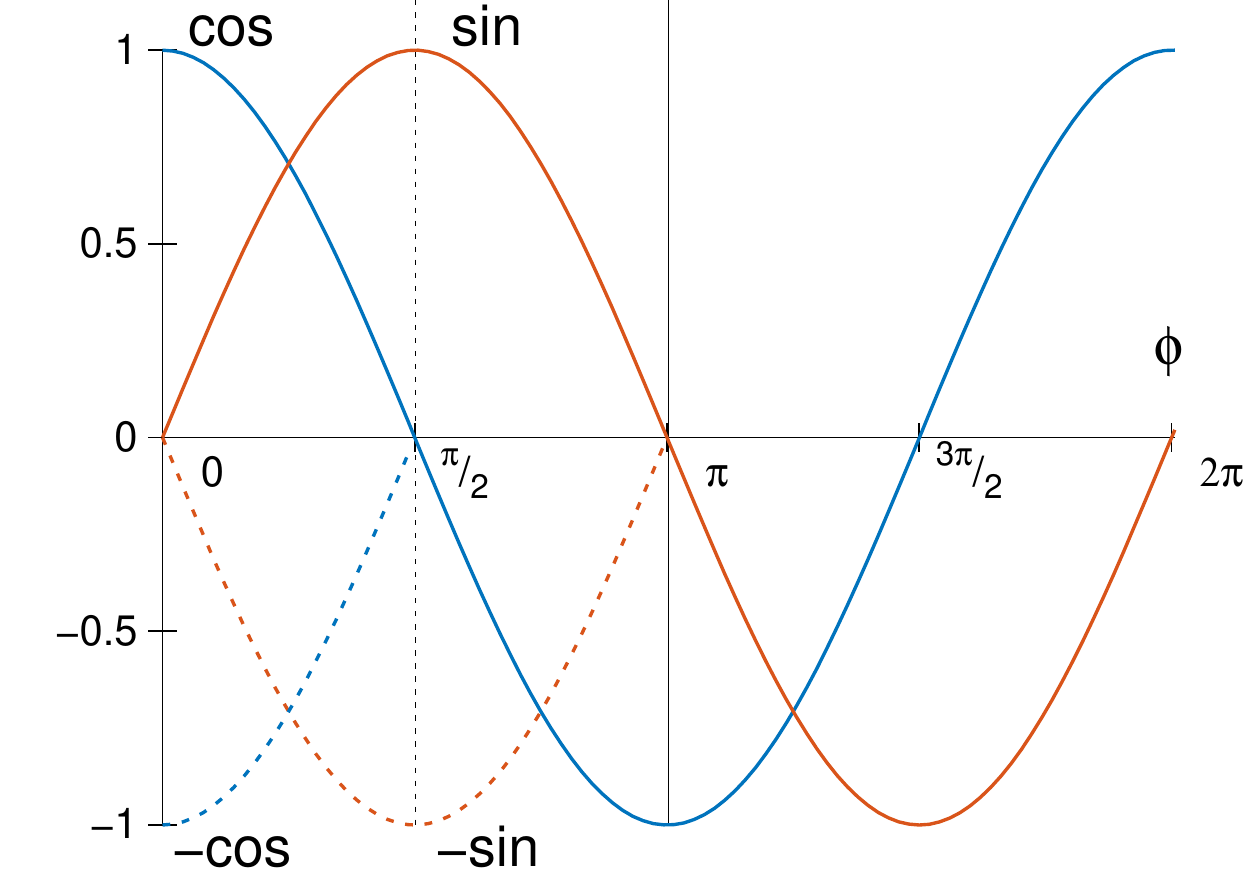}}
  \subfloat[][]{\includegraphics[width=0.33\textwidth]{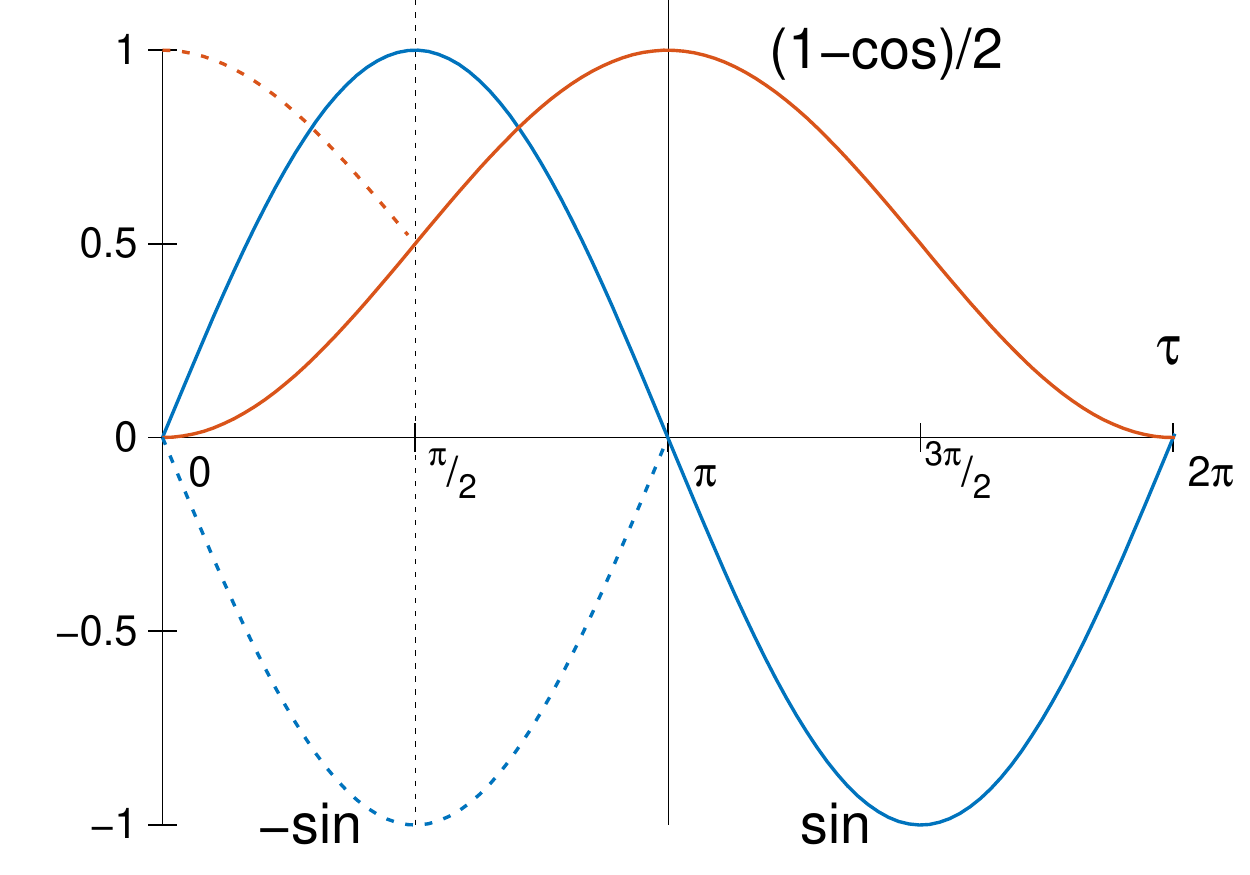}}
  \caption{Domain folding: (a) for the $\theta$ variable by
    Eq.~\eqref{eq:theta_ising_rule}, (b) for the $\phi$ variable by
    Eq.~\eqref{eq:phi_ising_rule_2}, and (c) for the amplitude by
    Eq.~\eqref{eq:tau_ising_rule_2}. New branches are dashed lines.}
  \label{fig:folding}
\end{figure*}
The difficulty in locating the global minimum in either \gls{QMF} or
\gls{QCC} theories can be rationalized as follows: Despite the
polylinear form of the energy functional, each trigonomitric factor is
a non-linear function with at least one extremum in the domain of
definition; ``individual'' extrema multiply as the number of variables
grows, and odds for locating the global minimum are greatly diminish.

As the local extrema of individual trigonometric factors are partially
responsible for this problem, we confine them in reduced domains by
creating multiple branches of the trigonomitric functions with the aid
of auxiliary discrete variables. We refer to this procedure as
``folding'' and illustrate it below for each class of continuous
variables separately.

Consider $\{\theta_i\}$ variables first. In the domain of definition,
Eq.~\eqref{eq:theta_domain}, each $\sin(\theta_i)$ has a maximum at
$\theta_i = \pi/2$ and is symmetric with respect to this line, while
$\cos(\theta_i)$ are monotonic and anti-symmetric. If we reflect a
piece of the cosine function on $[\pi/2, \pi]$ back to $[0, \pi/2)$,
we obtain a second branch which is negation of the original cos
function in the same range, see Fig.~\ref{fig:folding}a. Both branches
can be encoded in the reduced domain by new discrete variables $Z_i$
as:
\begin{equation}
  \label{eq:theta_ising_rule}
  \begin{array}{cl}
    \sin\theta_i \to &    \sin\theta_i, \\
    \cos\theta_i \to & Z_i\cos\theta_i, \\
  \end{array}
\end{equation}
where
\begin{equation}
  \label{eq:theta_ising_domain}
  \theta_i \in [0, \pi/2), \quad Z_i \in \{\pm 1\}
\end{equation}

For $\{\phi_i\}$ angles we have the same trigonometric functions as
for $\theta_i$-s, but on the $[0, 2\pi)$ domain [see
Eq.~\eqref{eq:phi_domain}]. This suggests that the domain folding can
be performed twice. Indeed, first we notice that $\cos(\phi)$ is even,
while $\sin(\phi)$ is odd with respect to the line $\phi = \pi$, see
Fig.~\ref{fig:folding}b. Thus, introducing new discrete variables
$Q_i$, we can write:
\begin{equation}
  \label{eq:phi_ising_rule_1}
  \begin{array}{rl}
    \sin\phi_i \to & Q_i\sin\phi_i, \\
    \cos\phi_i \to &    \cos\phi_i.
  \end{array}
\end{equation}
In the new domain, $\phi_i \in [0, \pi)$, the cosine function is odd,
but both branches of $\sin$ are even. Therefore, we can perform
another folding by introducing new discrete variables $W_i$:
\begin{equation}
  \label{eq:phi_ising_rule_2}
  \begin{array}{rl}
    \sin\phi_i \to & Q_i\sin\phi_i, \\
    \cos\phi_i \to & W_i\cos\phi_i,
  \end{array}
\end{equation}
where
\begin{equation}
  \label{eq:phi_ising_domain}
  \phi_i \in [0, \pi/2), \quad Q_i,W_i \in \{\pm 1\}.
\end{equation}

Amplitudes $\{\tau_i\}$ enter the \gls{QCC} energy expression as
$\sin(\tau_i)$ or $[1 - \cos(\tau_i)]$ functions
(Fig.~\ref{fig:folding}c). The domain folding can be performed twice:
first, with respect to the line $\tau = \pi$, which maps
$[1 - \cos(\tau)]$ to itself and creates two branches of $\sin(\tau)$,
\begin{equation}
  \label{eq:tau_ising_rule_1}
  \begin{array}{rl}
    \sin \tau_i \to & F_i\sin \tau_i, \\
    \left[1 - \cos \tau_i\right] \to & [1 - \cos \tau_i], \\
  \end{array}
\end{equation}
with
\begin{equation}
  \label{eq:tau_ising_domain_1}
  \tau_i \in [0, \pi), \quad F_i \in \{\pm 1\},
\end{equation}
and second, with respect to the line $\tau = \pi/2$, which creates
additional branch for $[1 - \cos(\tau)]$:
\begin{equation}
  \label{eq:tau_ising_rule_2}
  \begin{array}{rl}
    \sin \tau_i \to & F_i\sin \tau_i, \\
    \left[1 - \cos \tau_i\right] \to & [1 - G_i\cos \tau_i], \\
  \end{array}
\end{equation}
with
\begin{equation}
  \label{eq:tau_ising_domain_2}
  \tau_i \in [0, \pi/2), \quad F_i,G_i \in \{\pm 1\}.
\end{equation}

Note that after the foldings all branches of the trigonometric
functions become monotonic. Minimization of the \gls{QCC} energy
expression now requires continuous optimization over reduced domains
plus discrete optimization over $\{Z_i,\ Q_i,\ W_i, \ F_i,\ G_i\}$
variables. This mixed discrete-continuous optimization is done in two
alternating steps: 1) for fixed values of discrete variables the
continuous variables are optimized, 2) for fixed values of the
continuous variables the discrete variables are optimized. For
efficient discrete optimization the folded \gls{QCC} energy function
is expressed in the generalized Ising form
\begin{equation}
  \label{eq:genIsi}
  \hat H_\text{Is}^\text{gen} = \sum_i A_i {\hat z}_i + \sum_{ij} B_{ij}{\hat z}_i{\hat z}_j +
  \sum_{ijk} C_{ijk} {\hat z}_i {\hat z}_j {\hat z}_k + \dots
\end{equation}
where a single $\hat z_i$ operator represents one of the discrete
variables $\{Z_i,\ Q_i,\ W_i, \ F_i,\ G_i\}$, and coefficients $A_i$,
$B_{ij}$, and $C_{ijk}$ are derived from values of trigonometric
factors with fixed continuous variables. Obtaining the lowest
eigenstate of $\hat H_\text{Is}^\text{gen}$ is equivalent to the
discrete optimization step.

\subsection{Solving the generalized Ising Hamiltonians for various
  foldings}
\label{sec:gener-ising-hamilt}

Multiple levels of folding have been introduced in
Sec.~\ref{sec:domain-folding}: it is possible to fold once in
$\theta_i$ and twice in $\phi_i$ and $\tau_i$. While it is tempting to
use the maximum possible folding, there is a trade-off between
simplification of the energy landscape due to the domain reduction and
the complexity of the resulting Ising Hamiltonians. Each level of
folding~\footnote{The folding introduced by
  Eq.~\eqref{eq:tau_ising_rule_2} is special: each of
  $(1 - \cos \tau_i)$ factors gives rise to \emph{two} new terms, one
  of those is dependent of $G_i$ while the other is not. Overall, this
  leads to a $2^{N_\text{ent}}$-fold increase in the number of terms
  of the resulting Ising Hamiltonian. This increase, however, is
  moderate as compared to the size of the entire \gls{QCC} Ansatz
  ($3^{N_\text{ent}}$). On the other hand, a somewhat unexpected
  consequence of this additional expansion is: even at
  $\boldsymbol \tau = 0$ there are \emph{multiple} energy
  operators---one of those is the original $\hat H$, but others have
  the form
  \begin{equation*}
    \label{eq:extra_ops}
    \hat P_{k} \cdots \hat P_1 \hat H \hat P_1 \cdots \hat P_k, \quad k
    = 1, \dots, N_\text{ent}.
  \end{equation*}
  They appear each time when one of $G_i = -1$. It is not clear,
  however, if any lower than the \gls{QMF} energy values may come from
  these expressions---we left this question open for future studies. }
introduces additional $\hat z_i$ variables into Eq.~\eqref{eq:genIsi}.
Unfortunately, practical quantum annealers, like the D-Wave 2000Q
system, can \emph{not} deal with the generalized
form~\eqref{eq:genIsi}. To convert Eq.~\eqref{eq:genIsi} to a 2-local
form containing at most quadratic terms [Eq.~\eqref{eq:Ising_H}], one
has to introduce auxiliary variables (\emph{e.g.}
${\hat {z}_{ij}^{(2)}} = \hat z_i \hat z_j$) to lower the rank of
high-order terms and the corresponding constraints to avoid spurious
solutions. This step additionally increases the qubit count of the
discrete optimization. Therefore, calculations done on D-Wave's 2000Q
quantum annealer do not use the full folding scheme.

To assess capability of our folding technique in full, we simulate an
idealized quantum annealer on a classical computer by evaluating the
ground state of a generalized Ising Hamiltonian~\eqref{eq:genIsi}
using a direct diagonalization in the full multi-qubit Hilbert space
of the problem. Due to exponential growth of this space with the
number of qubits we treat only relatively small systems by this
``ideal Ising machine.''

We introduce the following notation to discuss performance of the
folding procedure at intermediate levels:
\begin{equation}
  \label{eq:folding_not}
  (m, n), \text{ where } 1 \le m \le 3, \ 0 \le n \le 2.
\end{equation}
$m$ indicates how many times the folding was done in mean-field
variables $\theta$ and $\phi$, while $n$ the number of foldings in
$\tau$. In particular, $m = 1$ means that the folding is done once in
$\theta$ variable for each qubit by Eq.~\eqref{eq:theta_ising_rule},
while $m = 2, 3$ means that $\theta$-folding is made once, but
additionally, $\phi$-foldings are made once or twice by
Eqs.~\eqref{eq:phi_ising_rule_1} or \eqref{eq:phi_ising_rule_2},
respectively. Overall, this introduces $N_q$, $2N_q$, or $3N_q$
$\hat z_i$ operators to the generalized Ising
Hamiltonian~\eqref{eq:genIsi} for $m = 1{-}3$. Additionally, since
there are no amplitudes the \gls{QMF} method, $n$ values may be
omitted to give a notation ``$(m,)$.''

For the \gls{QCC} method single- [$n=1$,
Eq.~\eqref{eq:tau_ising_rule_1}], double- [$n=2$,
Eq.~\eqref{eq:tau_ising_rule_2}], as well as no-folding ($n=0$)
variants are possible. As a result, $0$, $N_\text{ent}$, or
$2N_\text{ent}$ new operators can be defined for $n = 0, 1$, and 2,
respectively. Overall, there are $(m N_q + n N_\text{ent})$ $\hat z_i$
operators at the folding level $(m, n)$.

\section{Numerical studies}
\label{sec:numerical-studies}

\subsection{Preparatory calculations and optimization setup}
\label{sec:prep-calc-optim}

\begin{figure}
  \centering %
  \includegraphics[width=0.5\textwidth]{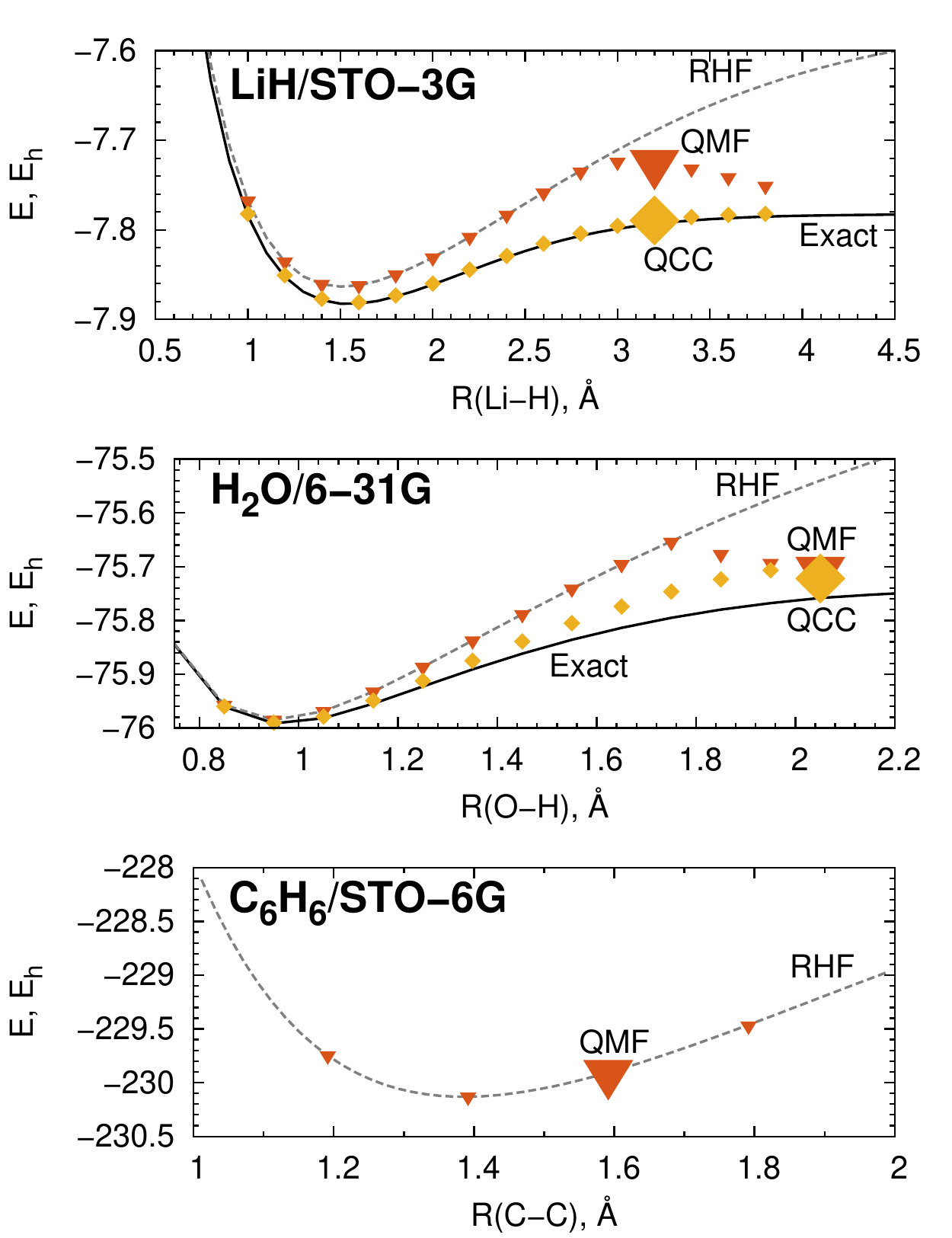}
  \caption{\Acrlong{PES} cuts for all molecules considered in the
    study. \Acrfull{RHF} and exact, \acrfull{CASCI}, curves were
    calculated on a classical computer using the \textsc{GAMESS}
    quantum chemistry package~\cite{gamessus-2}. Magnified symbols
    correspond to molecular configurations for which the domain
    folding plus annealing assessment is done.}
  \label{fig:pes}
\end{figure}

We calculated potential energy curves using the \gls{QMF} and
\gls{QCC} methods for the \ce{LiH} and \ce{H_2O} molecules, and the
\gls{QMF} method for \ce{C_6H_6} on both a classical computer and the
D-Wave 2000Q system. Near equilibrium geometries the \gls{QCC} method
provides the chemical accuracy, $\le \SI{1}{\kilo\cal\per\mol}$,
within a chosen basis and active space. However, to make our examples
more challenging, we consider a few molecular structures outside the
equilibrium, namely: $R(\ce{Li-H}) = \SI{3.20}{\angstrom}$ (the
equilibrium value is \emph{ca.} \SI{1.54}{\angstrom}), the
symmetrically stretched to $R(\ce{O-H}) = \SI{2.05}{\angstrom}$ water
molecule (the equilibrium value is \emph{ca.} \SI{0.96}{\angstrom}),
and a symmetrically elongated benzene ring with
$R(\ce{C-C}) = \SI{1.5914}{\angstrom}$ (the equilibrium value is
\emph{ca.} \SI{1.34}{\angstrom}), see Fig.~\ref{fig:pes}.
For more technical details on preparatory calculations see
Table~\ref{tab:mol_prop}.

The domain folding technique paired with a quantum annealer is
assessed against a simple gradient-based local optimization (several
popular local gradient-based continuous optimization algorithms are
compared in Appendix~\ref{sec:optim-meth}). Namely, starting from a
random guess for Bloch angles and amplitudes and using a local
optimization algorithm with and without annealing, we collect
statistics how often each of the minima has been reached out of 100
runs.

Generalized Ising Hamiltonians for different folding levels were
generated as described in Sec.~\ref{sec:gener-ising-hamilt}. Annealing
is done on the ideal Ising machine and the D-Wave 2000Q system, but
for the latter the corresponding generalized Ising Hamiltonians were
converted to a 2-local Ising form~\eqref{eq:Ising_H} with the aid of
D-Wave's Ocean software~\cite{dwave-ocean}. Biases ($h_i$) and
coupling terms ($J_{ij}$) of the Ising Hamiltonian~\eqref{eq:Ising_H}
were computed on a classical computer; the resulting Hamiltonians were
embedded onto D-Wave's 2000Q using the minor-miner
algorithm~\cite{minorminer} using a cutoff of \num{1e-2}. Qubit counts
for each embedding are reported in Table~\ref{tab:dwave-embedding}. A
constant annealing time of \SI{100}{\micro\second} was used for all
molecules, and the number of samples was equal to 1000. Bloch angles
and amplitudes were updated using L-BFGS-B gradient optimization
algorithm based on the minimum energy sampled from the annealer.
\begin{table}
  \centering
  \caption{The number of qubits used by D-Wave's 2000Q system to
    represent a \gls{QMF}/\gls{QCC} problem for a given molecule.
    Ranges reflect variation of this count due to neglecting the small
    terms in parametrized Ising Hamiltonians uploaded onto the
    annealer.}
  \begin{tabularx}{1.0\linewidth}{@{}XYY@{}}
    \toprule
    Molecule    & \multicolumn{2}{c}{Qubit count in the Ising form} \\ \cmidrule{2-3}
                & (1,0)~folding       & (1,1)~folding \\\midrule
    \ce{LiH}    & 9--14               & 28-35 \\
    \ce{H_2O}   & 14--22              & 128--136 \\
    \ce{C_6H_6} & 800--900\footnotemark[1]          & -- \\
    \bottomrule
  \end{tabularx}
  \footnotetext[1]{Only \gls{QMF} simulations were performed, so that
    in the absence of amplitudes the folding scheme is $(1,)$.}
  \label{tab:dwave-embedding}
\end{table}

\subsection{\gls{QMF} and \gls{QCC} simulations for LiH}
\label{sec:lih-qmf-qcc-simulations}

\begin{figure}
  \centering %
  \captionsetup[subfigure]{labelformat=empty}
  \subfloat[][]{
    \label{subfig:lih-qmf} %
    \includegraphics[width=0.5\textwidth]{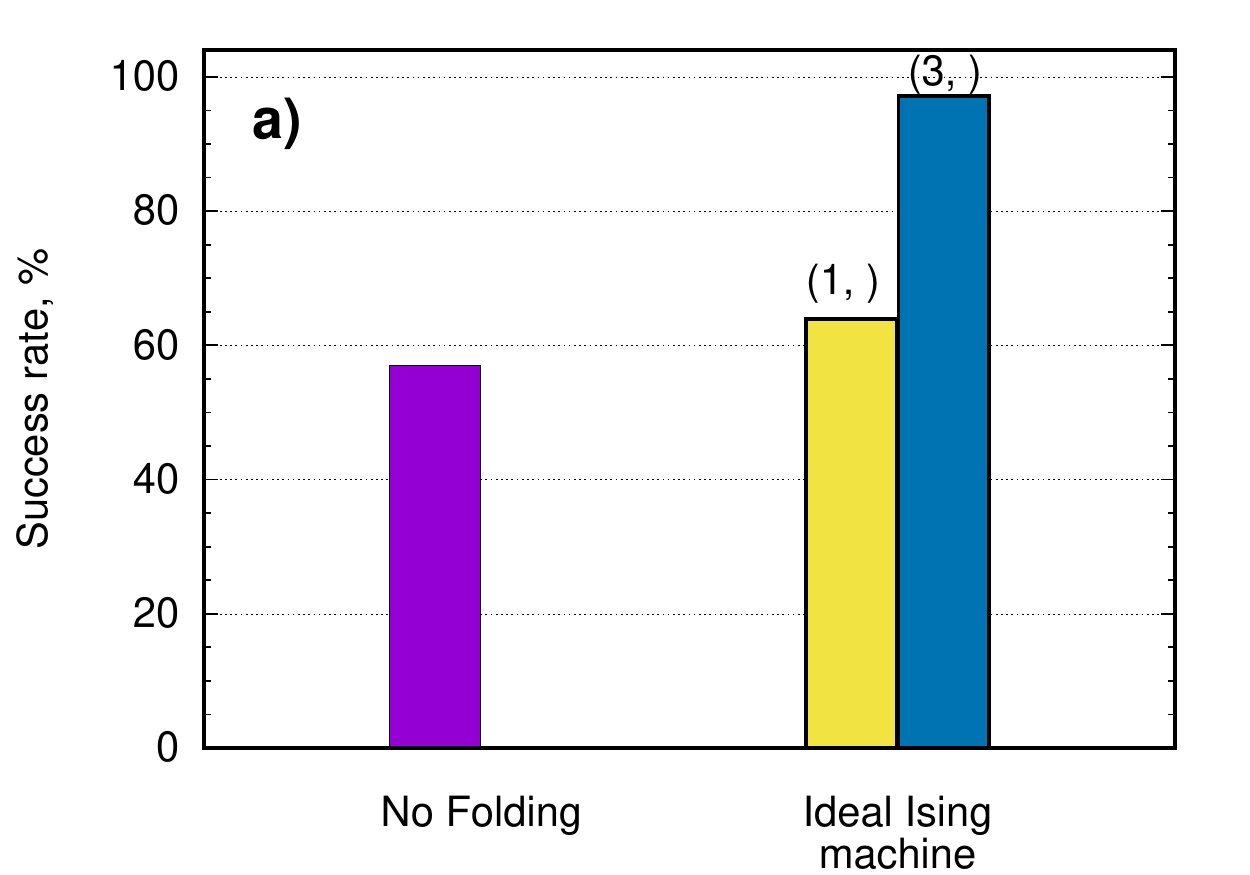}}
  
  \subfloat[][]{
    \label{subfig:lih-qcc}
    \includegraphics[width=0.5\textwidth]{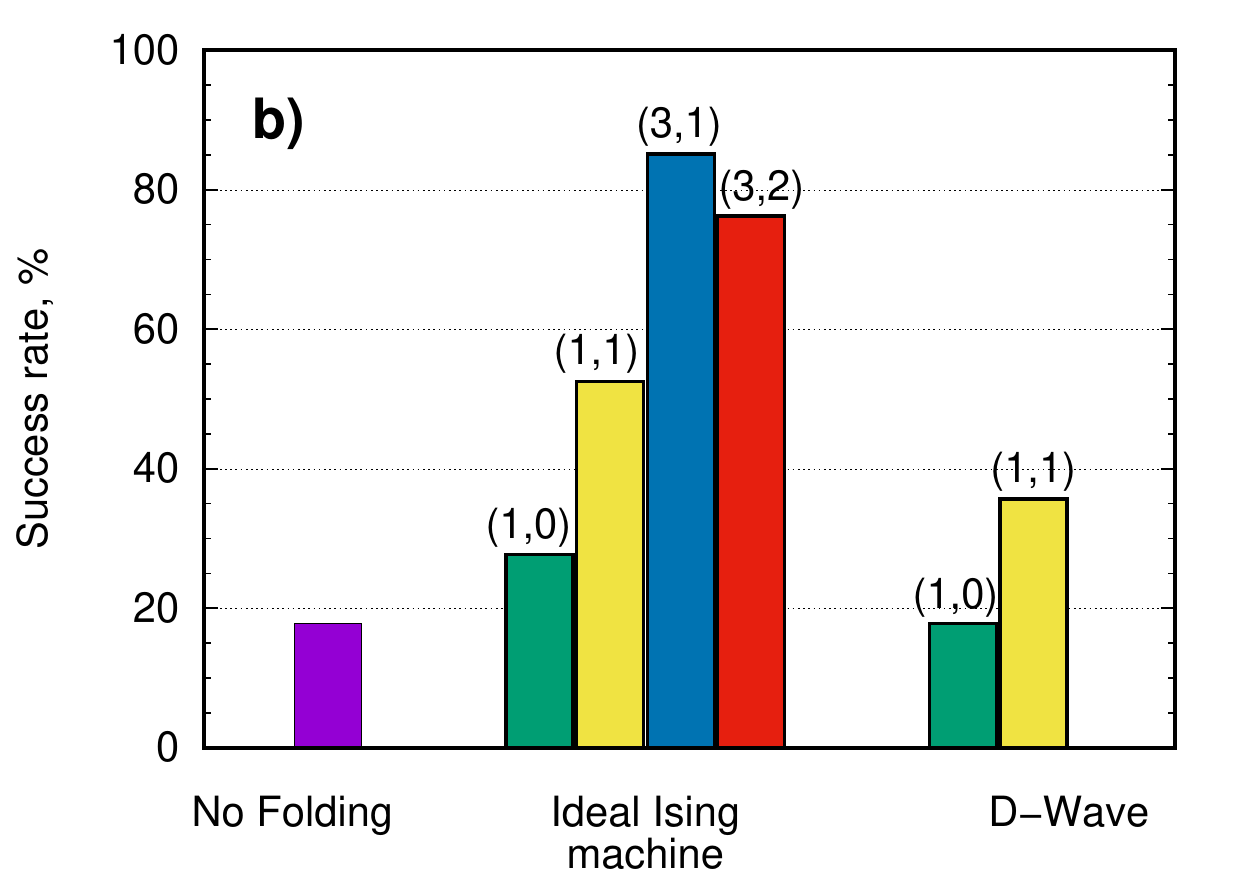}}
  \caption{Global minimum search success rate for the
    \gls{QMF}~\protect\subref{subfig:lih-qmf} and
    \gls{QCC}~\protect\subref{subfig:lih-qcc} calculations of the
    stretched LiH molecule. Different folding levels are labelled
    according to Sec.~\ref{sec:gener-ising-hamilt}.}
  \label{fig:lih}
\end{figure}

A stretched \ce{LiH} molecule is an example of a spin-broken system.
Although the molecular orbitals that have been used to generate the
qubit Hamiltonian~\eqref{eq:spin_ham} were taken from the \gls{RHF}
calculations, the intrinsic ability of the \gls{QMF} method to break
symmetry~\cite{Ryabinkin:2019/jctc/249} lead to the solution that was
not a pure singlet, much like as in the \emph{unrestricted}
Hartree--Fock method, see Fig.~\ref{fig:pes}. In such a situation one
would expect a competition between several low-lying symmetry-broken
solutions which are characterized by markedly different values of
Bloch angles. Indeed, even in the \gls{QMF} method without the domain
folding there are several low-lying local minima, which lower a chance
to reach the ground state as can be seen in Fig.~\ref{subfig:lih-qmf}.
Successive domain foldings in Bloch angles eliminate chances to
converge to anything but the true ground state almost entirely. Thus,
the domain folding technique greatly enhances probability to reach the
true minimum even in a presence of symmetry breaking. It must be noted
that symmetry breaking does not necessarily imply incorrect
calculations; instead, such a phenomenon indicates existence of a
complicated open-shell ground state in molecules with partly ruptured
bonds or containing transition metal atoms.

Introducing electron correlation at the \gls{QCC} level does change
statistics, but not the physics. For highly stretched molecules the
primary role of electron correlation is symmetry restoration
\emph{via} proper mixing of open-shell atomic states that emerge in a
course of dissociation. Since symmetry constraints are quite rigid, we
do not expect multiple low-lying minima. Indeed, \gls{QCC}
calculations bring the total energy to be within
\SI{2}{\kilo\cal\per\mol} from exact, but the main effect of amplitude
folding is already seen at the $(1,1)$ level, see
Fig.~\ref{subfig:lih-qcc}. While for the unfolded problem the
probability of finding the true ground state is merely
\SI{20}{\percent}, it raises to almost \SI{60}{\percent} and, finally,
to more than \SI{85}{\percent} by going to $(1,1)$ and $(3,1)$ levels,
respectively. It is also clear that the Bloch angle foldings are more
important than the amplitude foldings, as it should be for the
symmetry-controlled case: moving to the highest $(3,2)$ level,
\emph{i.e.} simplifying the amplitude optimization even more, does not
improve the probability to reach the ground state.

Fig.~\ref{subfig:lih-qcc} also assesses the performance of the domain
folding technique on the D-Wave 2000Q system. It is clear that the
D-Wave 2000Q annealer, similar to the ideal Ising machine,
systematically improves the chance to reach the ground state as the
level of folding increases. There is still a minor gap in efficiency
between the real annealer and an idealized device; we comment on the
possible reasons more in Sec.~\ref{sec:water}.

\subsection{\gls{QCC} simulations for H$_2$O}
\label{sec:water}

\begin{figure}
  \centering %
  \includegraphics[width=0.5\textwidth]{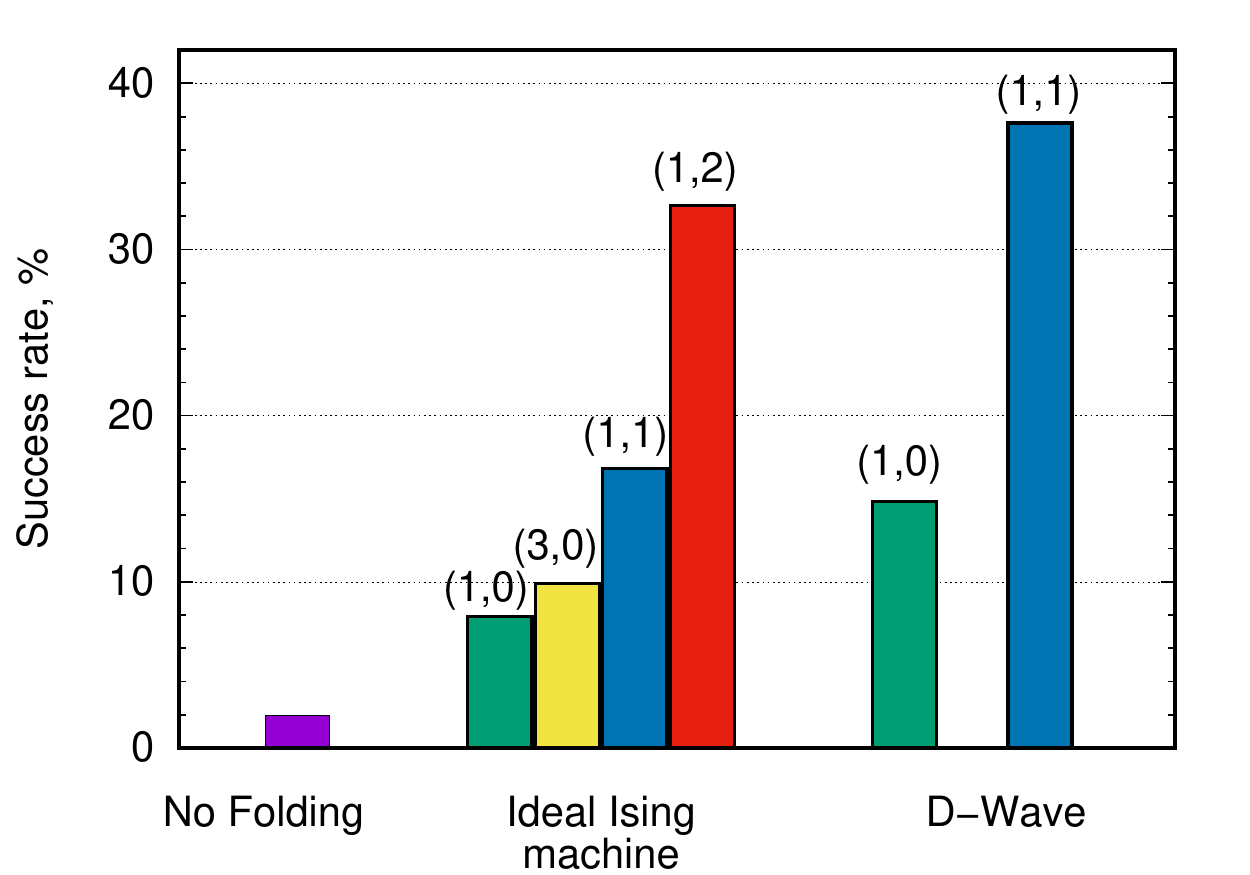}
  \caption{Global minimum search success rate for the \gls{QCC}
    calculations of the stretched \ce{H2O} molecule.}
  \label{fig:h2o_qcc}
\end{figure}
A symmetric stretch of the \ce{H2O} molecule is the classical test for
methods aimed at treating the strong correlation
problem~\cite{Abrams:2005/cpl/284}. The most difficult situation,
however, is not when bonds are completely broken but rather when they
are ``half-broken''. In such a nuclear configuration atomic states are
heavily mixed with relative weights that are controlled not only by
symmetry but also the interaction strength.

As evident from Fig.~\ref{fig:h2o_qcc}, the \gls{QCC} optimization
problem is extremely difficult: the probability to find the ground
state without folding is just a few percent---one needs several dozens
of tries to reach the ground state once! Various folding schemes shown
in Fig.~\ref{fig:h2o_qcc} do improve the situation, albeit to a
different extent. Contrary to the \ce{LiH} case, Bloch-angle foldings
help only moderately, increasing the chance to reach the ground state
to $\sim \SI{10}{\percent}$ at the $(3,0)$ level. This essentially
means that the correlated solution is of the same symmetry as the
mean-field one, and the mean-field solution is not symmetry-broken.
This is additionally corroborated by the fact that even a single
amplitude folding raises the chance to reach the ground state to
$\SI{17}{\percent}$, which is a substantial improvement over the
unfolded result and any of the Bloch-folded counterparts. Finally,
$(1,2)$ folding brings the chance to $\SI{33}{\percent}$: now one
needs only 3 runs on average to reach the ground state.

To our surprise, the D-Wave 2000Q system performs \emph{better} then
an ideal Ising machine, as can be seen from Fig.~\ref{fig:h2o_qcc}.
There are several plausible explanations of this phenomenon: different
efficiency of local optimization methods (\gls{MMA} on the idealized
Ising machine \emph{vs.} L-BFGS-B for the D-Wave 2000Q system, see
Appendix~\ref{sec:optim-meth}) or smoothing of the energy landscape on
the D-Wave system because of neglecting small terms in Ising
Hamiltonians, so that the shallow minima are missed. We ruled out the
smoothing as a reason: implementing similar strategy on the ideal
Ising machine shown that the statistics is either almost unaffected
(for small cutoffs) or so strongly skewed that is clearly incorrect.
Thus, it is likely that slight variations of the efficiency of
optimization algorithms plus statistical fluctuations may explain the
difference. However, these variations do not affect the main
conclusion: the domain folding provides \emph{systematic} improvement
of the global search.

\subsection{\gls{QMF} simulations for C$_6$H$_6$}
\label{sec:c6h6}

Our final example is large-scale, 36-qubit, \gls{QMF} calculations of
a slightly expanded \ce{C6H6} ring. In this case it is not possible to
simulate the problem on an ideal Ising machine due to excessive memory
requirements.

Results show that a single $\theta$ folding drastically increases
chances to obtain the global minimum of energy: from less than
\SI{5}{\percent} to \SI{40}{\percent}. It might be somewhat surprising
that \gls{QMF} calculations without the folding face so profound
difficulties in reaching the true minimum near the equilibrium
geometry, where a single-determinant solution is stable to any kind of
symmetry breaking. Moreover, such calculations have to be equivalent
to the \gls{RHF} ones (see Fig.~\ref{fig:pes}), which can be performed
with ease on a classical computer. The resolution of this apparent
paradox lies in the fact that, as was already noted at the end of
Sec.~\ref{sec:brief-descr-qmf}, the \gls{QMF} method has no access to
quantities like orbital energies and, has to search through multiple
orbital population patterns to find the lowest-energy one. At the same
time, in the \gls{RHF} method with the Fock matrix diagonalization,
the Aufbau principle~\cite{Saunders:1973/ijqc/699} immediately rules
out the majority of high-energy minima.

We provide in-depth analysis of the problem in
Appendix~\ref{sec:equiv-qmf-rhf}; here we only emphasize that a single
domain folding in $\theta_i$ variables plus quantum annealing can be
perceived as a substitute of the Aufbau principle for the \gls{QMF}
method.

\section{Conclusions}
\label{sec:conclusions}

We have presented and assessed the domain folding technique that
allows one to exploit optimization capacity of quantum annealers in
quantum chemistry calculations. The domain folding uses multiple local
symmetries that present in the \acrfull{QMF} and \acrfull{QCC}
methods---which were originally formulated for use on universal
quantum computers---to greatly enhance chances to reach the global
minimum of electronic energy. On the set of electronic structure
calculations for \ce{LiH}, \ce{H2O}, and \ce{C6H6} molecules we have
demonstrated the advantages of the domain folding both on the ideal
Ising machine and on the D-Wave 2000Q system.

Although preparatory stages for \gls{QMF}/\gls{QCC} calculations
require substantial classical precomputations, the use of annealers
makes the \gls{QMF} and \gls{QCC} methods quite promising in
situations with symmetry breaking/strong correlation, in which
multiple solutions of the electronic structure problem
exist~\cite{Celestino:2003/mp/1937}. To the best of our knowledge, the
satisfactory treatment for such cases were not available before, but
with the domain folding technique coupled to quantum annealers, they
can be treated much better.

Additionally, we found (see Sec.~\ref{sec:c6h6}) that any variational
qubit Ansatz, as a global minimization problem, would experience
difficulties in locating the global minimum (even in the simplest
parametrization, like our \gls{QMF} form), due to an exponential
number of energy minima. Recently, such difficulties have been
reported for another \gls{VQE} approach in
Ref.~\citenum{Lee:2019/jctc/311}. The domain folding technique
mitigates the problem, and for the case of \gls{QMF} can be thought of
as a substitute for the missing Aufbau principle, see
Appendix~\ref{sec:equiv-qmf-rhf}. Moreover, the domain folding
technique for the \gls{QCC} method, which goes beyond the mean-field
treatment, may open opportunities to study strongly correlated
systems.

\begin{acknowledgments}
  A.F.I. acknowledges financial support from the Natural Sciences and
  Engineering Research Council of Canada.
\end{acknowledgments}

\appendix

\glsresetall

\section{Preparatory calculations}
\label{sec:prep-calc}

Essential details on preparatory calculations are given in
Table~\ref{tab:mol_prop}.
\begin{table*}
  \centering
  \caption{Fermionic and qubit Hamiltonians construction parameters
    for molecules used in the work. Canonical set of the Hartree--Fock
    molecular orbitals is used throughout.}
  \begin{tabularx}{1.0\textwidth}{>{}Xccc@{}}
    \toprule
    Property                 & \multicolumn{3}{c}{Molecule} \\ \cmidrule{2-4}
                             & \ce{LiH}\footnotemark[1]               & \ce{H_2O}\footnotemark[1]  & \ce{C6H6}   \\ \midrule
    Molecular configuration  & $R(\ce{Li-H}) = \SI{3.20}{\angstrom}$  & $ R(\ce{O-H})    = \SI{2.05}{\angstrom}$ & $R(\ce{C-C}) = \SI{1.5914}{\angstrom}$ \\
                             &                                        & $\angle \ce{HOH} = \SI{107.6}{\degree}$  & $R(\ce{C-H}) = \SI{1.0802}{\angstrom}$ \\
    Atomic basis set\footnotemark[2] & STO-3G                         & 6-31G                                    & STO-6G \\
    Total number of molecular orbitals & 6                                      & 13                                       & 36     \\
    Number of spin-orbitals
    in active space, $N_\text{so}$ & 6                                 & 8                                        & 36 \\
    Number of electrons
    in active space          & 2                                      & 4                                        & 18 \\
    Fock space dimension, $2^{N_\text{so}}$ & 64                        & 256                                      & \num[round-mode=figures,round-precision=3,scientific-notation=true]{6.872e10} \\
    Fermion-to-qubit mapping & parity\footnotemark[3]                 & \acrfull{BK} &  \acrfull{BK} \\
    Qubit count, $N_q$       &  4\footnotemark[4]                                     & 6\footnotemark[4]                                        & 36 \\
    Entanglers for \gls{QCC}, $\hat P_i$ & $\hat x_2 \hat y_0,\ \hat x_3 \hat x_2 \hat x_1 \hat y_0,\ \hat y_3 \hat x_2 \hat y_1 \hat y_0$
                             & $\hat P,\, \hat z_5 \hat z_2 \hat P,\ \hat x_5 \hat x_2 \hat  P$ & \\
                             &          & $\hat x_2 \hat y_0,\ \hat y_5 \hat x_3 ,\ \hat x_4 \hat y_1$ & -- \\
                             &          & $\hat  P = \hat x_4 \hat x_3 \hat x_1 \hat y_0$ & \\

    \bottomrule
  \end{tabularx}
  \footnotetext[1]{More details on the electronic structure
    calculations and \gls{QMF}/\gls{QCC} setup can be found in
    Ref.~\citenum{Ryabinkin:2018/jctc/6317}.} \footnotetext[2]{From
    the Basis Set Exchange library~\cite{emsl-2}.}
  \footnotetext[3]{Described in
    Ref.~\citenum{Nielsen:2005/scholar_text}.}
  \label{tab:mol_prop}
  \footnotetext[4]{Qubit parity symmetries have been used to reduce
    the qubit count by 2 as compared to $N_\text{so}$.}
\end{table*}

\section{Comparison of different continuous optimization methods}
\label{sec:optim-meth}

\begin{figure}
  \centering
  \includegraphics[width=0.5\textwidth]{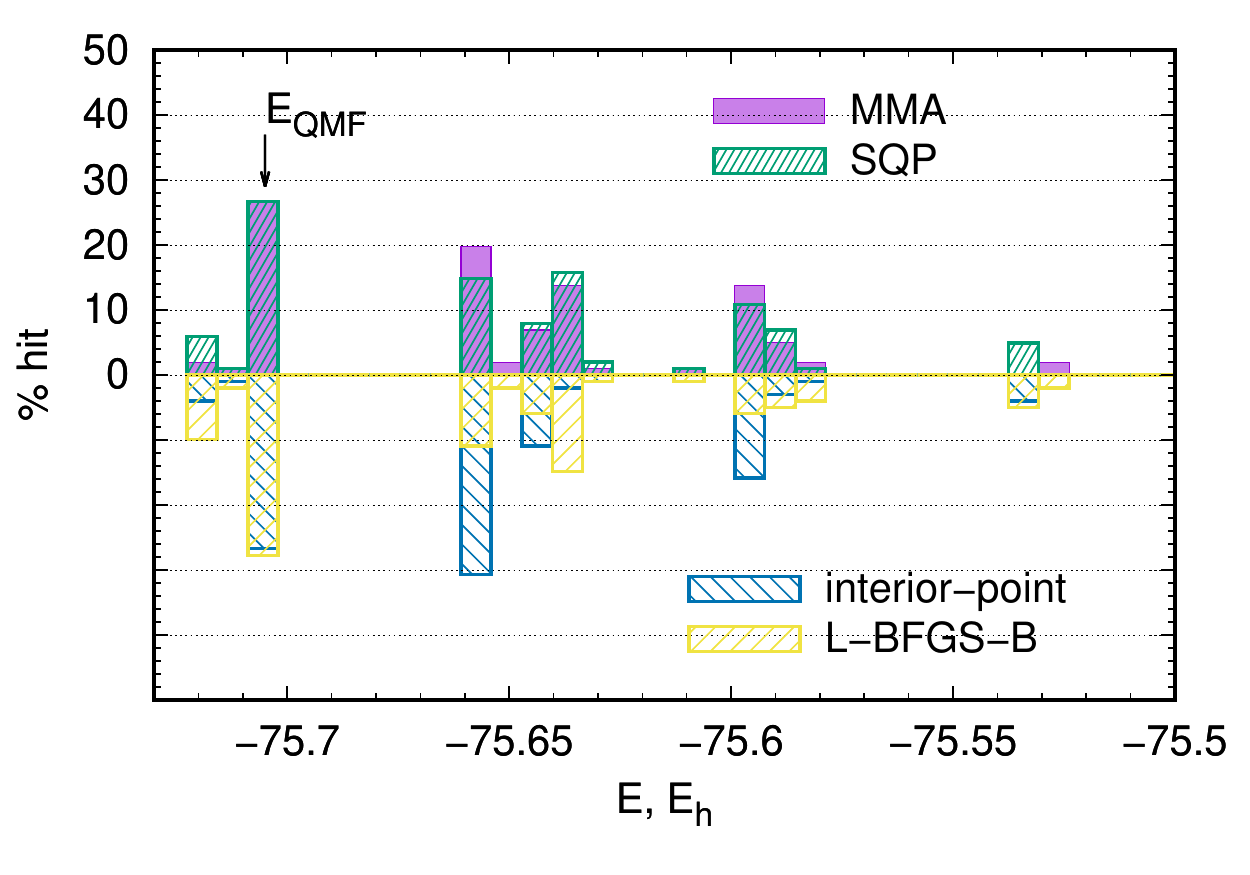}
  \caption{Stationary point location statistics for different local
    optimization algorithms in the \gls{QCC} method for the \ce{H2O}
    molecule. A half of the results are mirrored to improve
    readability. Acronyms are explained in
    Appendix~\ref{sec:optim-meth}.}
  \label{fig:h2o_no_annealer}
\end{figure}
We tested several local continuous optimization methods on a single
problem of \gls{QCC} energy optimization for the \ce{H_2O} molecule to
confirm that neither of them is superior to others in global search.
These methods are:
\begin{enumerate}
  \setlength{\itemsep}{0.5ex} %

\item The \gls{MMA}~\cite{MMA-Svanberg:2002/jopt/555} as implemented
  in Ref.~\citenum{NLopt-Johnson:2011} available through the
  \texttt{NLopt.jl} Julia\cite{Julialang} package. It is a globally
  convergent method, which is especially convenient when
  function/gradient evaluations are very time-consuming, which is the
  case for an idealized Ising machine.

\item The L-BFGS-B method~\cite{lbfgs-scipy-Byrd:1995/jsc/1190,
    lbfgs-scipy-Zhu:1997/atms/550} available through Python API on
  D-Wave's system. It is less robust than \gls{MMA}, but the cost of
  discrete energy minimization is determined by the D-Wave 2000Q
  system.

\item[3--4.] The \gls{SQP}~\cite{Nocedal:2006} and
  interior-point~\cite{interior-point-Byrd:1999/jopt,
    interior-point-Byrd:2000/mp/149, interior-point-Waltz:2006/mp/391}
  methods from the \textsc{matlab}~\cite{MATLAB:2018a} suit. These
  methods were not used in calculations with the domain folding,
  rather, they sre included as references to the readily available
  algorithms that may be chosen to solve \gls{QMF} or \gls{QCC}
  problems on a classical computer.
\end{enumerate}

Assessment of different continuous optimizers in the case of the
\ce{H2O} molecule is presented in Fig.~\ref{fig:h2o_no_annealer}.
Clearly all methods are quite comparable in their ability to find
various local minima in the system, and for all of them finding the
global minimum is indeed a difficult task. It is interesting, that
less robust methods, such as L-BFGS-B and interior-point ones, are
slightly more efficient in the global search problem. It is likely
that these methods may simply miss some of the high-energy shallow
minima, thus improving the statistics for deeper minima including the
ground state.

\section{On the equivalence of \gls{QMF} and \gls{RHF} calculations}
\label{sec:equiv-qmf-rhf}

For many molecules, especially near their equilibrium configurations,
the \gls{RHF} single-determinant wave function is a qualitatively
correct representation of the exact one; it is stable with respect all
possible spin and electron-number variations. The \gls{QMF}
calculations that start from the fermionic
Hamiltonian~\eqref{eq:qe_ham} written in the basis of the canonical
Hartree--Fock orbitals have to converge to the \gls{RHF}
energy~\cite{Ryabinkin:2018/jcp/214105}. However, the pristine
\gls{QMF} energy minimization has non-negligible probability to
converge to other, higher-energy solutions, as evident from
Fig.~\ref{subfig:lih-qmf} and results of Sec.~\ref{sec:c6h6}. Here we
investigate this \gls{QMF} deficiency, provide the reasoning why the
\gls{RHF} method is largely immune to it, and find out why the domain
folding in $\theta_i$ variables plus quantum annealing can be
perceived as a cure.

First, we observe that a single-determinant wave function $\ket{\Phi}$
can be characterized not only by the total number of electrons, $N_e$,
but also the orbital population vector, in which every orbital is
either populated (1) or not (0), according to the mean values of
orbital population operators
\begin{align}
  \label{eq:ni-ops}
  \hat n_i = {} & a_i^\dagger a_i, \ i = 1, \dots, N_\text{so}, \\
  \label{eq:ni-means}
  \braket{\hat n_i} = {} & \braket{\Phi | \hat n_i | \Phi} = \{0, 1\}. 
\end{align}

After any of the fermion-to-qubit transformations, the
operators~\eqref{eq:ni-ops} acquire the Ising
form~\cite{Seeley:2012/jcp/224109}:
\begin{equation}
  \label{eq:ni_qubit}
  \hat n_i \to \hat n_i(\hat z_i).
\end{equation}
In turn, Eq.~\eqref{eq:ni_qubit} becomes a function of only
$\cos(\theta_i)$ after application of Eq.~\eqref{eq:qmf_rule}. A
simple observation shows that the integer values in
Eq.~\eqref{eq:ni-means} are, in general, compatible with only
$\theta_i = 0$ or $\pi$, when $\cos\theta_i = \pm 1$. This essentially
means that the \gls{RHF} solution is characterized by \emph{fixed}
$\theta_i$ angles either 0 or $\pi$; which, in turn, implies that
continuous optimization of $\theta_i$ is not necessary. Moreover, as
follows from Eq.~\eqref{eq:coh_state}, for $\theta_i = 0$ or $\pi$ the
wave function does not depend (apart from an irrelevant global phase)
on $\phi$ angles too. Thus, the \gls{QMF} energy optimization is
reduced to the discrete part only. However, without the annealing it
is still exponentially hard, which explains why the \gls{QMF} method
has difficulties in locating the global minimum.

It must be noted that such a problem does not impact strongly common
implementations of the \gls{RHF} method on a classical computer. Most
of the computational schemes repeatedly construct and diagonalize the
Fock matrix, which gives access to orbital energy estimates. The
latter can be sorted, and only lowest $2N_e$ of them (``the Aufbau
principle'') can be used to built the ground electronic wave function.
In general, for a typical molecular system without stretched bonds or
transition-metal ions, only a small fraction of orbitals have energies
close to the Fermi level and can be potentially swapped during the
self-consistent cycle, which translates into the small number of
potential candidates for the global minimum. In the opposite case,
however, when many orbitals are in energy proximity to the Fermi
level, the traditional Hartree--Fock algorithms may experience
difficulties in locating the lowest-energy solution. In such
situations the \gls{QMF} method with domain folding and quantum
annealing may become the method of choice, as it can deal with the
plethora of local minima efficiently.

\end{document}